\documentclass[aps,prl,showpacs,superscriptaddress,reprint]{revtex4}
\usepackage{amsmath,amssymb,amsfonts,color,graphicx,bm,amsthm,tabularx,datetime,array}

\newcolumntype{M}[1]{>{\centering\arraybackslash}m{#1}}

\newcommand{\D}{$\mathcal{D}~$}
\newcommand{\C}{$\mathcal{C}~$}
\newcommand{\V}{$\mathcal{V}_{max}~$}

\begin{document}

\date{\today~~\currenttime}

\title{Transient to Zero-Lag Synchronization in Excitable Networks}

\author{H. Brama}
\affiliation{Gonda Interdisciplinary Brain Research Center and the Goodman Faculty of Life Sciences, Bar-Ilan University, Ramat-Gan 52900, Israel}

\author{Y. Peleg}
\affiliation{Department of Physics, Bar-Ilan University, Ramat-Gan 52900, Israel}

\author {W. Kinzel}
\affiliation{Institute for Theoretical Physics, University of W\"{u}rzburg, Am Hubland, 97074 W\"{u}rzburg, Germany}

\author{I. Kanter}
\affiliation{Gonda Interdisciplinary Brain Research Center and the Goodman Faculty of Life Sciences, Bar-Ilan University, Ramat-Gan 52900, Israel}
\affiliation{Department of Physics, Bar-Ilan University, Ramat-Gan 52900, Israel}

\begin{abstract}
The scaling of transient times to zero-lag synchronization in networks composed of excitable units is shown to be governed by three features of the graph representing the network:
the longest path between pairs of neurons (diameter), the largest loop (circumference) and the loop with the maximal average out degree.
The upper bound of transient times can vary between $O(1)$ and $O(N^2)$, where $N$ is the size of the network, and its scaling can be predicted in many scenarios from finite time accumulated information of the transient.
Results challenge the assumption that functionality of neural networks might depend solely upon the synchronized repeated activation such as zero-lag synchronization.
\end{abstract}

\maketitle

The cooperative behavior of time delay networks composed of excitable units is of fundamental interest in many fields of science and especially in neuroscience where the building blocks of the brain, the neurons, function as threshold units.
Psychological and physiological considerations entail that formation and functionality of neuronal cell assemblies depend upon the reverberation modes such as zero-lag synchronization (ZLS)
\cite{brain 1,brain 2,brain 3,brain 4}.
Several mechanisms for the emergence of this phenomenon have been suggested, including the global network quantity - the greatest common divisor (GCD) of neuronal circuit delay loops.
This nonlocal phenomenon, where neurons which do not have common drive are synchronized, pinpoints the interplay between the network topology and the number of reverberating clusters, where neurons belonging to a cluster are in ZLS \cite{GCD 1, GCD 2}.
This interplay was recently verified in an experimental procedure on a circuit of neurons embedded within a large-scale network of cortical cells in-vitro \cite{in vitro} as well as in experiments of phase locking in time-delayed coupled laser networks and networks of mutually coupled chaotic lasers \cite{Optics Express, Nixon}.

The synchronous activity of excitable networks initiated by limited classes of external drives is nowadays well understood.
Nevertheless, little is known about the transient times to achieve the synchronous activity.
Since neural phenomena often occur on very short time scales, there seems to be an inherent temporal gap between the relatively possibly long transient trajectory to steady state pattern, and the postulate that functionality of the network is based on reverberating patterns only.
In case transient times to reverberating modes are too long compared to the time scale of neural phenomena, an inference of accumulated information along the trajectory might occur in the brain activity.
Furthermore, from theoretical point of view it is intriguing to know whether the synchronous mode of activity and the transient time are determined by the same graphical features and whether networks with the same steady state patterns have similar transient times.
In this Letter we show that the scaling of the transient time to ZLS of excitable network is governed by the diameter, the circumference and the loop with the maximal average out degree.
The upper bound of the transient to ZLS can vary between $O(1)$ and $O(N^2)$, where $N$ is the size of the network.

\begin{figure}[b]
 \includegraphics[width=0.5\textwidth]{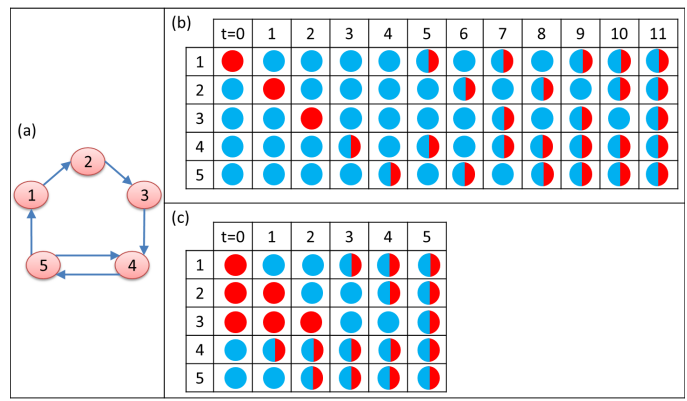}
   \caption{
  The transient time to ZLS is exemplified by the mixing argument \cite{GCD 1}
  for a network consists  of 5 neurons.
  (a) Schematic of a homogeneous network where all delays are equal to $\tau$.
  (b) An example where the initial stimulation (t=0) is given to the neuron labeled by 1.
  Each one of the neurons mixes the colors of its driving neurons from the previous time step, e.g. neuron 4 mixes the colors of neurons 3 and 5 from the previous time step.
  Red (dark) color indicates an evoked spike - an excited neuron.
  The transient time is equal to 11 in units of $\tau$.
  (c) An example where initial stimuli are simultaneously given to neurons 1,2 and 3.
  The transient time to ZLS is equal to 5.
  }
 \label{fig.zls}
\end{figure}

\begin{table*}
\begin{center}
{
\begin{tabular}{M{0.03\textwidth}||M{0.5\textwidth}||M{0.05\textwidth}|M{0.05\textwidth}|M{0.1\textwidth}|M{0.05\textwidth}}
    & Network              & $O(\mathcal{D})$ & $O(\mathcal{C})$ & $O(\mathcal{V}_{max})$  & O(T)\\ \hline
    &                                                               &     &     &           &       \\
  1 & \includegraphics[height=1.74cm]{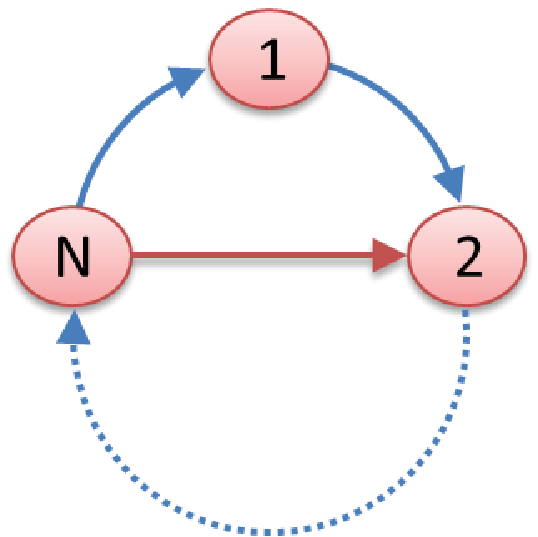}                      & $N$ & $N$& $ N^{-1}$  & $N^2$ \\ \hline
    &                                                               &     &     &           &       \\
  2 & \includegraphics[height=1.74cm]{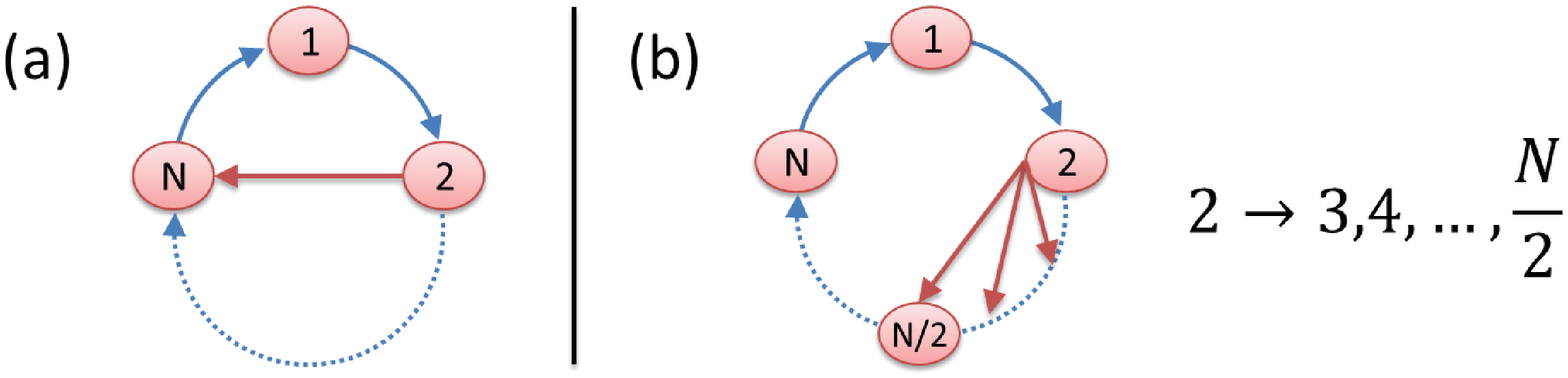}                      & $N$ & $N$& $1$        & $N$   \\ \hline
    &                                                               &     &     &           &       \\
  3 & \includegraphics[height=1.74cm]{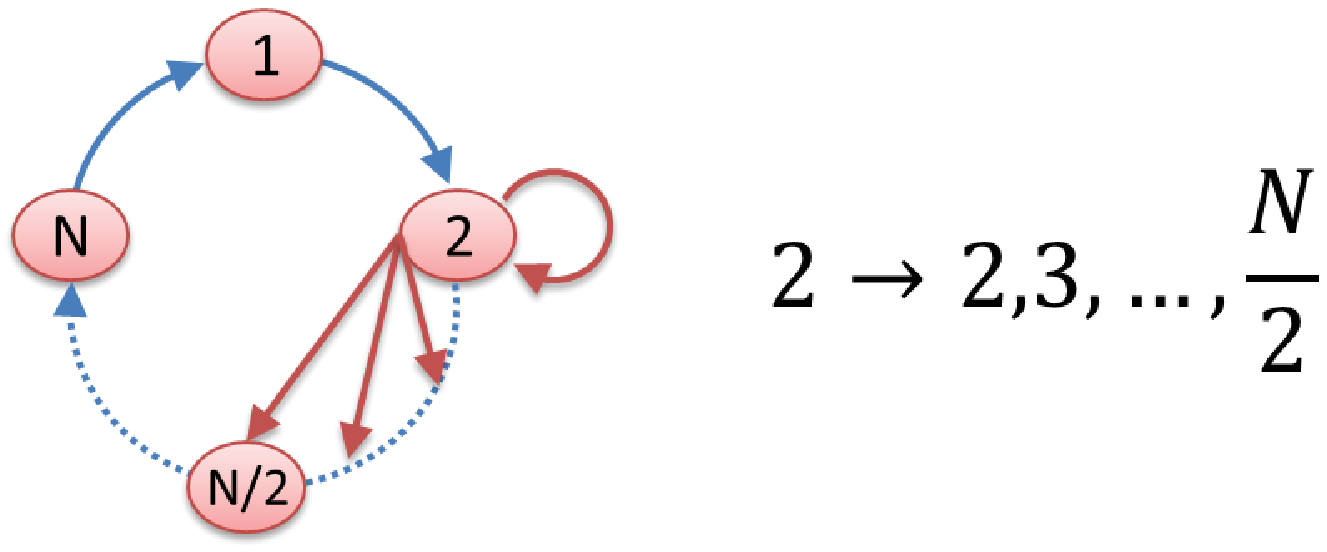}                      & $N$ & $N$& $N$        & $N$   \\ \hline
    &                                                               &     &     &           &       \\
  4 & \includegraphics[width=0.5\textwidth,height=1.74cm]{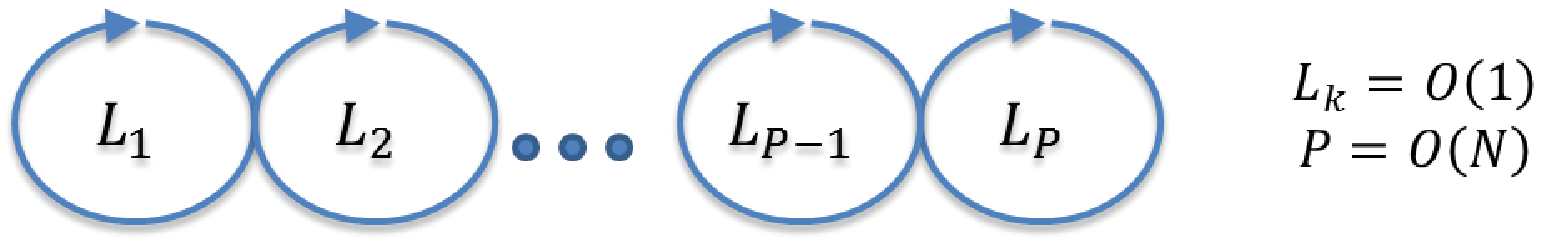}  & $N$ & $1$& $1$        & $N$   \\ \hline
    &                                                               &     &     &           &       \\
  5 & \includegraphics[height=1.74cm]{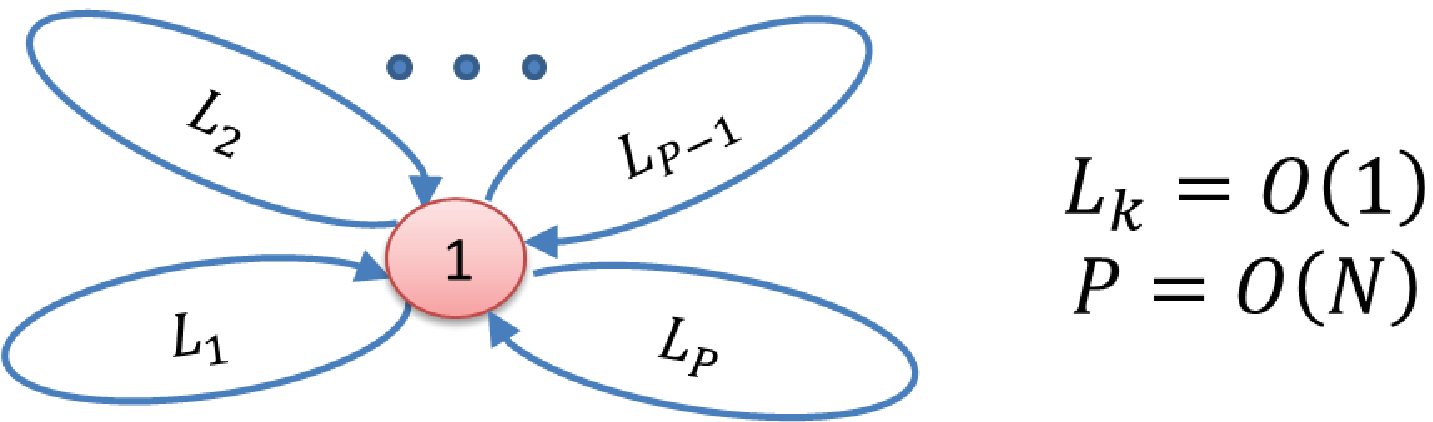}                      & $1$ & $1$& $N$        & $1$   \\ \hline
    &                                                               &     &     &           &       \\
  6 & \includegraphics[height=1.74cm]{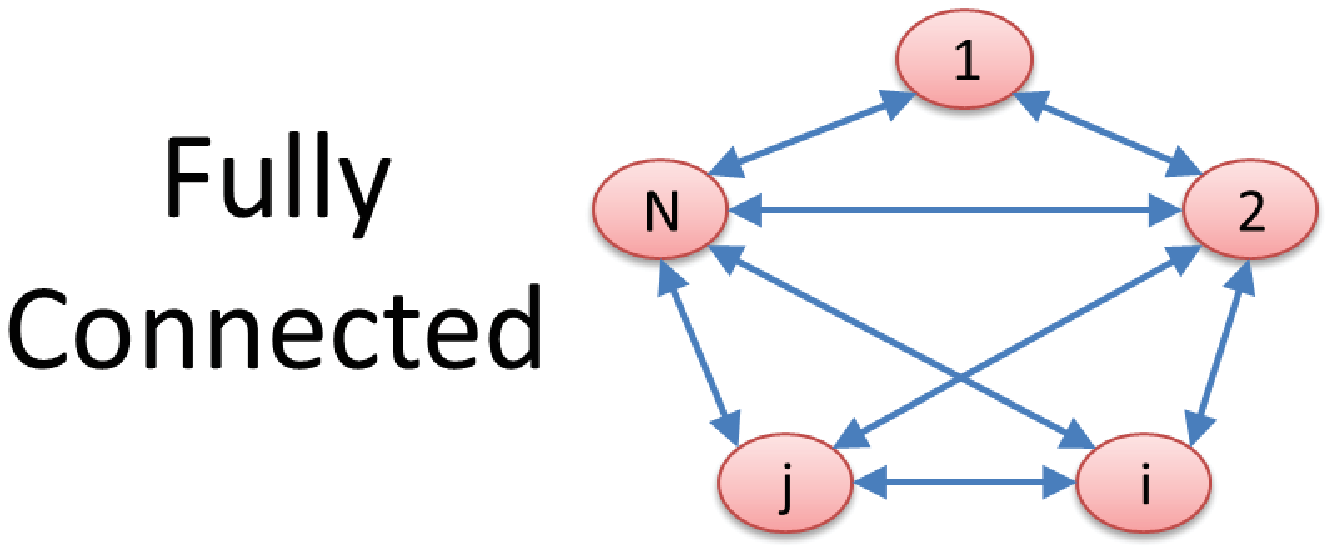}                      & $1$ & $N$& $N$        & $1$   \\ \hline
\end{tabular}
}
\end{center}
\caption{
Illustration of the transient time, $T$, for networks leading to ZLS (second column) where \D and \C scale either as $O(1)$ or as $O(N)$.
Rows (1-3) exemplify networks with \D and \C of $O(N)$ but different scaling for \V.
Rows (4-6) exemplify networks where the transient is dominated by \D.
}
\label{tab:gt}
\end{table*}

The examined networks consist of $N$ boolean neurons denoted by
$\bm{X}(t),(X_i=\{0,1\},~i=1,2,\ldots N)$,
where $X_i=1$ stands for an evoked spike from the i\textit{-th} neuron.
The dynamics of a network with homogeneous time-delays, $\tau$, can be described as an $N$ dimensional iterative map
\begin{equation}\label{eq: update}
    \bm{X}\left(t+1\right)=sgn\left(M \bm{X}\left(t\right)\right)=sgn\left(M^t \bm{X}\left(0\right)\right)
\end{equation}
where time steps are normalized by $\tau$, $sgn(\cdot)$ is the signum function and $M$ is the adjacency matrix of the graph representing the network, $M_{i,j}=1$ iff there is a connection from neuron $j$ to neuron $i$, otherwise $M_{i,j}=0$.
We assume that $M$ represents a strongly connected graph, i.e. there is a legal path among any pair of neurons, and the GCD of the network loops is equal to 1.

The transient to ZLS is defined as the minimal time steps until all neurons evoked spikes simultaneously, $\bm{X}(t)=\bm{1}$ and it may vary among different initial stimuli, Fig. \ref{fig.zls}.
Nevertheless, we denote by $T$ the maximal transient time over all possible initial stimuli (given simultaneously to a selected subset of neurons).
Quantitatively, $T$ is defined such that all elements of $M^T$ are positive, $sgn(M^T)=1$.
Such matrices are called primitive matrices and $T$ is known to be bounded by \cite{exponent 1,exponent 3}
\begin{equation}\label{eq. bound}
T \le \left(N-1\right)^2+1=O\left(N^2\right).
\end{equation}
This upper bound transient time, $T$, is taken over all possible networks of size $N$ and can be far above the actual transient, e.g. Fig. \ref{fig.zls}(b).

We propose that the upper bound scaling of $T$ given by
\begin{equation}\label{eq: transient}
T =  O\left(\frac{\mathcal{C}}{\mathcal{V}_{max}} + \mathcal{D}\right)
\end{equation}
where \C and \D stand for the circumference (the largest loop) and the diameter (the longest path among all the shortest paths between pairs of neurons), respectively,
\begin{equation}\label{eq: V}
    \mathcal{V}_{L}   =  \frac{1}{\|L\|} \sum_{i\in L} \left(\mathrm{out~degree}\left(X_i\right) -1\right)
\end{equation}
indicates the average number of generated new spikes per time-delay $\tau$ while a spike is traveling along the loop $L$ (of length $\|L\|$) and
\begin{equation}\label{eq: V max}
    \mathcal{V}_{max}=\displaystyle\max_{L\in loops} \mathcal{V}_{L}
\end{equation}
is the maximal generating rate of new spikes along a single loop in the network.
The scaling, Eq. \ref{eq: transient}, is based on the necessary condition $T \ge \mathcal{D}$ since the initial stimulation must span over the entire network in order to achieve ZLS.
On the other hand, by definition of the diameter, during number of \D steps the initial stimulation goes through every possible neuron of the network.
As a consequence, after \D steps the initial evoked spike occupies simultaneously each loop and the filling progress toward ZLS is performed in parallel in the entire network loops.
For T>\D, the rate of generating new spikes is at least \V, Eq. \eqref{eq: V max}, hence the time to fill the largest loop, \C, is at most $O(\mathcal{C}/\mathcal{V}_{max})$.
Now within \D additional steps each spike occupies simultaneously each network loop and ZLS is achieved, which establishes the scaling - Eq. \eqref{eq: transient}.

To exemplify the scaling, Eq. \eqref{eq: transient}, we first restrict the discussion to networks where \C and \D are of $O(1)$ or $O(N)$ and \V can take one of the following orders $O(N),O(1)$ or $O(N^{-1})$ as illustrated by Table \ref{tab:gt}.
The scaling of the transient time for each network was calculated using Eq. \eqref{eq: transient} and has been confirmed in simulations.

Case (1) of Table \ref{tab:gt}, consists of two loops of length $O(N)$ generating only one new spike per cycle, $\mathcal{V}_{max}=\frac{O(1)}{O(N)}=O(N^{-1})$. Hence, the transient time is dominated by the first term of Eq. \eqref{eq: transient} and scales as $T=O(N^2)$, saturating the upper bound scaling for $T$, Eq. \eqref{eq. bound}. Reversing the direction of the red arrow in Table \ref{tab:gt}(1) results in a small loop of length $O(1)$, Table \ref{tab:gt}(2), characterized by \V$=O(1)$. Following Eq. \eqref{eq: transient}, $T$ scales now linearly with the system size.
Adding $O(N)$ outgoing connections to neuron 2 of Table \ref{tab:gt}(1), as illustrated in Table \ref{tab:gt}(2b), results in a \D$=O(N)$, although all loops of the network are still $O(N)$. Since $O(N)$ new spikes are generated at neuron 2, \V$=\frac{O(N)}{O(N)}=O(1)$ and the transient is $T=O(N)$.
Adding only a single self-connection to neuron 2 in the previous case, Table \ref{tab:gt}(3), the self-loop generates $O(N)$ new spikes at each time step and \V$=\frac{O(N)}{O(1)}=O(N)$.
However, an initial stimulus to neuron $N/2$, for example, travels \D$=O(N)$ time steps before arriving at the self-loop thus the transient is $T=O\left(\frac{1}{N}+N\right)=O(N)$.

Table \ref{tab:gt}(4) represents the case where \D$=O(N)$ but \C$=O(1)$ and each loop creates $O(1)$ new spikes, thus \V$=O(1)$. Nevertheless $T=O(N)$ as spikes must travel from one end of the network to the other end in order to achieve ZLS.
Connecting all the loops to a single neuron, Table \ref{tab:gt}(5), reduces the diameter, \D$=O(1)$, while each loop generates $O(N)$ new spikes. As a consequence, \V$=O(N)$ but as before the diameter determines the transient $T=O(\frac{1}{N}+1)=O(1)$.
A fully connected graph, Table \ref{tab:gt}(6), is very similar to the previous case but the circumference is $O(N)$, nevertheless the transient remains $T=O(\frac{N}{N}+1)=O(1)$.

\begin{figure}[t]
  \includegraphics[width=0.5\textwidth]{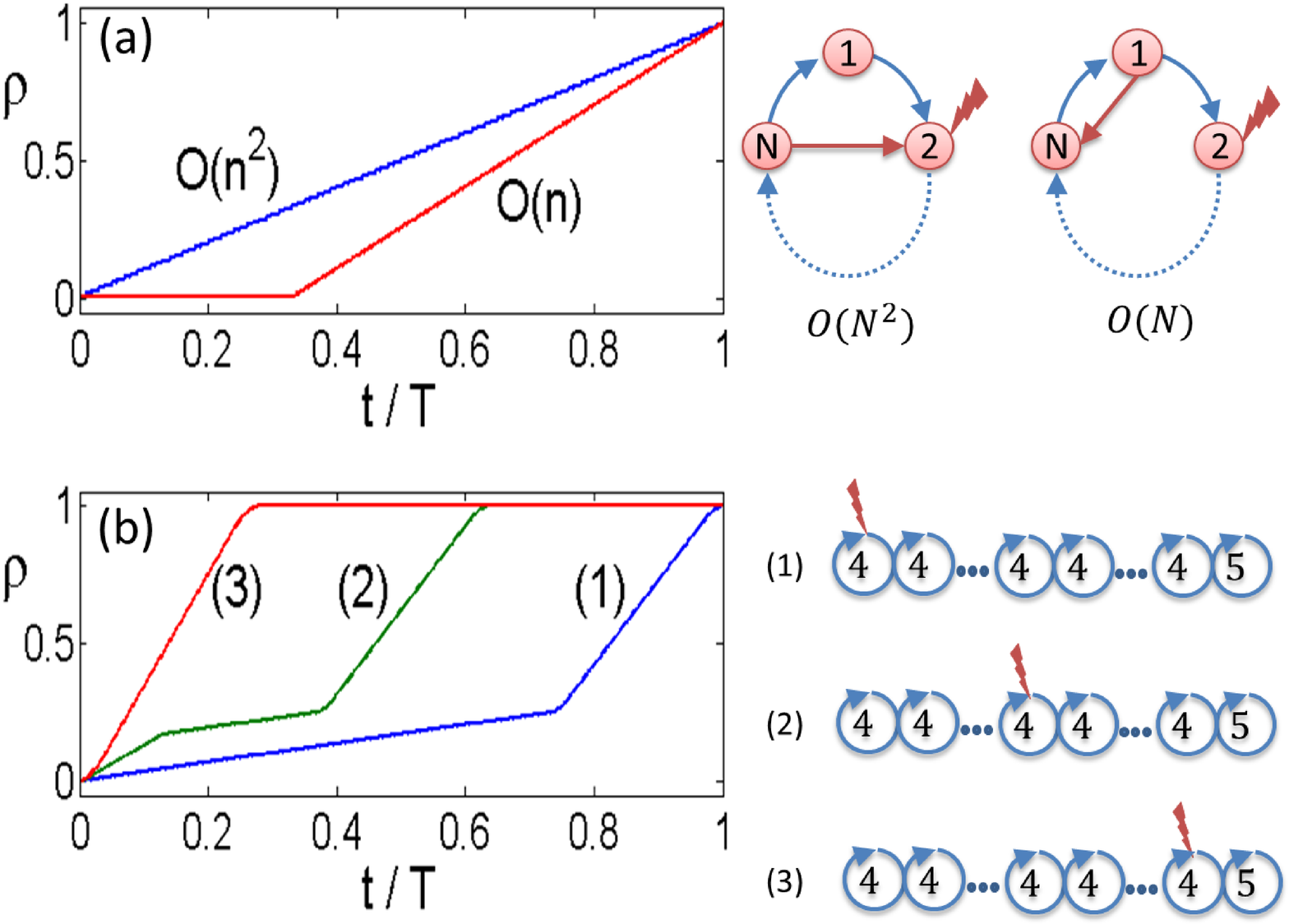}\\
   \caption{
  Fraction of neurons which are in ZLS as a function of time for various cases and initial conditions.
  The (red) lightning denotes the position of the initial stimulation to the network and numbers inside the loops denote their lengths.
  (a) Linear growth of $\rho(t)$ obtained in simulations with $N=101$ for the two networks shown on the right.
  For $T=O(N)$ there is an initial time, bounded by the diameter, where $\rho(t)=1/N$.
  (b) 100 loops of length 4 and the rightmost loop is of length 5.
  The shape of $\rho(t)$ and the transient time depend on the location of the stimulations.
  For case (1) it takes $O(\mathcal{D})$ time steps for a spike to arrive at the loop of length 5 which is essential to achieve ZLS.
  For initial conditions (2) and (3) this time is shorter and the time to ZLS is reduced.
}\label{fig2}
\end{figure}

Finding $\mathcal{V}_{max}$ might be an exhaustive computational task since it requires the inspection of all the network loops.
Moreover, the transient time and its scaling can be altered by the specific type of the stimuli.
Hence, we examine the question whether the scaling of the transient time can be predicted from finite steps of the transient.
Specifically, we define the quantity
\begin{equation}\label{eq: rho}
    \rho(t)=\frac{1}{N}\sum_{i=1}^{N} X_i(t)
\end{equation}
indicating the fraction of neurons which are in ZLS at time $t$.
Fig. \ref{fig2}(a) indicates that $\rho(t)$ grows linearly with time for a case where $T=O(N^2)$.
It is evident that the scaling of $T$ can be predicted from a partial segment of the transient, however the generality of this consequence is in question.
Fig. \ref{fig2}(a) exemplifies $\rho(t)$ for \ref{tab:gt}(2a) where $T=O(N)$ and for a finite fraction of the transient $\rho$ is constant.
This plateau is originated from an initial stimulation given at a distance proportional to \D from the junction (neuron 1, which is responsible to the production of new spikes).
Similar plateau can occur for $T=O\left(N^2\right)$, however it is negligible as the diameter is at most $O(N)$. Note that the scaling of $T$ can be predicted also for such cases from a partial segment of the transient with a nonzero slope of $\rho(t)$.

A more structured profile for $\rho(t)$ is depicted in Fig. \ref{fig2}(b) where $T$ is dominated by the the diameter and its shape is sensitive to the location of the initial stimulation.
The positive slope of $\rho(t)$ changes once in curve (1) and twice in curve (2) while curve (3) consists of only a single positive slope.
The network consist of loops of length 4 only, except for the rightmost one which is of length 5, which is essential for GCD=1.
For case (1), where the initial stimulation is given to the leftmost loop, the rightmost loop is first visited after $O\left(\mathrm{D}\right)$ time steps, but as the initial stimulation is moved to the right, case (2) and (3), this time is shortened.
Nevertheless, the scaling of $T$ can be predicted for all these examined cases from any finite segment with a positive slope of $\rho(t)$.
An exception class is networks with an out degree $O(N)$ of at least one neuron and $T=O(N)$, e.g. Table \ref{tab:gt}(2a,3).
In such cases, $\rho(t)$ is discontinuous as $O(N)$ spikes are created at once.

\begin{figure}[t]
  \includegraphics[width=0.5\textwidth]{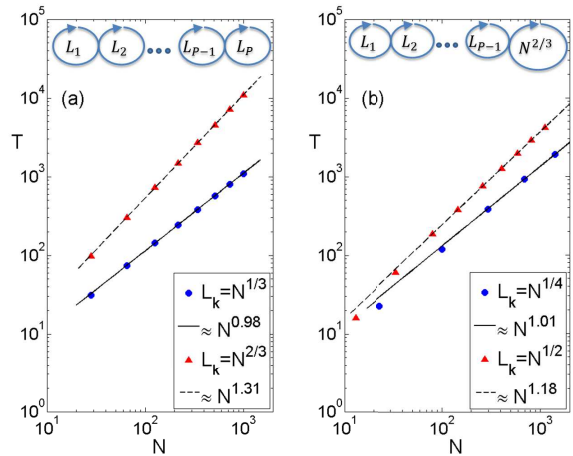}\\
   \caption{
  Extension of the scaling of the transient time, Eq. \eqref{eq: transient}, to the case where \C and $\mathcal{V}_{max}^{-1}$ are both $O\left(N^\alpha\right), 0<\alpha<1$.
  (a) The lengths of all loops scales as $O\left(N^\alpha\right)$, $\alpha=1/3$ (lower blue) and $\alpha=2/3$ (upper red).
  Simulation results indicates that for $\alpha=1/3$ $T\propto N^{0.98}\sim O(N)$, while for $\alpha=2/3$ $T\propto N^{1.31}\sim O(N^{4/3})$ which are both in a good agreement with the theoretical prediction, Eq. \eqref{eq: transient}.
  (b) A network where one of the loops scales differently, \C$=O(N^\beta),~0<\alpha<\beta<1$.
  The transient time scales as $T=O(N^{\beta+\alpha}+N)$.
  For $\alpha=1/4$ (lower blue) the estimated scaling is $T\propto N^{1.01}$ while for $\alpha=1/2$ the estimation is $T\propto N^{1.18}$, in a fairly good agreement with the scaling predicted by Eq. \eqref{eq: transient}.
}
\label{fig3}
\end{figure}

Hitherto, we have assumed that \C and \D are either $O(1)$ or $O(N)$. However, the scaling of the transient time, Eq. \eqref{eq: transient}, is valid also for $\mathcal{V}_{max}^{-1}$ and \C or \D of order $N^\alpha,~0<\alpha<1$.
Such a scenario is exemplified for case (4) of Table \ref{tab:gt} where loops are of the same order $L_k=O(N^\alpha)~,~(0<\alpha<1)$.
In such a case \C$=O\left(N^{\alpha}\right)$ and $\mathcal{V}_{max}=O\left(N^{-\alpha}\right)$.
Following Eq. \eqref{eq: transient}, the transient should scale as $T=O\left(N^{2\alpha} + N\right)$ since $\mathcal{D}=O(N)$.
For $\alpha<0.5$ the transient is dominated by the diameter, $T=O(N)$, while for $\alpha>0.5$ the transient is governed by \C and \V, $T=O\left(N^{2\alpha}\right)$.
Figure \ref{fig3}(a) presents simulation results with $N \le 2000$ for the two cases $(\alpha=1/3$ and $~2/3)$.
The  estimated scaling of the transient time is $T\propto N^{0.98} \sim O(N)$ and $T\propto N^{1.31} \sim O(N^{4/3})$ respectively, which is in a good agreement with the theoretical prediction, Eq. \eqref{eq: transient}.

For the case where one of the loops scales differently, $O(N^\beta)~(0<\alpha<\beta<1)$, the transient time should scale as  $T=O\left(N^{\alpha+\beta} + N\right)$ since the generating rate of new spikes remains $\mathcal{V}_{max}=O\left(N^{-\alpha}\right)$ but the circumference is now \C$=O\left(N^{\beta}\right)$.
A new scaling for the transient time is expected for $\alpha+\beta > 1$ where the length of the loops determine the scaling of the transient.
For the case $\alpha+\beta=1/4+2/3~<1$ the transient time is dominated by the diameter is estimated in simulations to scale as $T\propto N^{1.01}$, Fig. \ref{fig3}(b), while for $\alpha+\beta=1/2+2/3~>1$ the transient time is estimated to scale as $T\propto N^{1.18}$ which is close to the theoretical prediction of Eq. \eqref{eq: transient}, $T=O(N^{7/6})$.

In conclusion, the scaling law of the transient time to achieve ZLS is established as a function of only three properties of the network: the circumference, the diameter and the loop with maximal average out degree, where all these parameters can scale as $O(N^\alpha), 0 \le \alpha \le 1$.
The scaling law was found to be a tight  upper bound for many classes of networks but cannot be proven in general.
Moreover, we have shown that typically the transient can be predicted from a finite steps of the transient.
Finally we note that for a matrix $M$ where  the sum of each row is normalized to a constant, e.g. $\sum_{j}M_{i,j}=1$,  it has been shown that the stability of synchronization is determined by the eigenvalue gap around the largest eigenvalue \cite{swen}.
Similarly, the scaling of the transient time is inversely proportional to the eigenvalue gap.
However, in our dynamics ZLS is not an eigenvector of $M$, hence the transient time is not determined by the eigenvalue gap but rather by the graphical properties of the network, Eq. \eqref{eq: transient}.

\end{document}